%% file: ANA-EXOT-2019-35-PAPER.tex
\documentclass[PAPER, atlasdraft=false, cernpreprint=true, UKenglish, texmf, orcidlogo]{atlasdoc}
\usepackage[subcaption]{atlaspackage}
\usepackage{atlasbiblatex}

\usepackage{atlasphysics}
\graphicspath{{logos/}{figures/}}

\usepackage{ANA-EXOT-2019-35-PAPER-defs}

\usepackage{rotating}
\usepackage{adjustbox}

\usepackage{tabularx}
\usepackage{longtable}
\usepackage{ltablex}
\usepackage{verbatimbox}
\usepackage[normalem]{ulem}
\usepackage{url}
\usepackage{float}
\usepackage{placeins}
\usepackage[normalem]{ulem}
\usepackage{fancyvrb}
\AtlasTitle{Search for heavy neutral leptons in decays of $W$ bosons produced in 13~TeV $pp$ collisions using prompt signatures in the ATLAS detector}

\AtlasAbstract{%
The existence of right-handed neutrinos with Majorana masses below the electroweak scale could help address the origins of neutrino masses, the matter--antimatter asymmetry, and dark matter. In this paper, leptonic decays of $W$ bosons from 140~fb$^{-1}$ of 13~TeV proton--proton collisions at the LHC, reconstructed in the ATLAS experiment, are used to search for heavy neutral leptons produced through their mixing with muon or electron neutrinos in a scenario with lepton number violation. The search is conducted using prompt leptonic decay signatures. The considered final states require two same-charge leptons or three leptons, while vetoing three-lepton same-flavour topologies. No significant excess over the expected Standard Model backgrounds is found, leading to constraints on the heavy neutral lepton's mixing with muon and electron neutrinos for heavy-neutral-lepton masses. The analysis excludes $|U_{e}|^2$ values above $8\times 10^{-5}$ and $|U_{\mu}|^2$ values above $5.0 \times 10^{-5}$ in the full mass range of 8--65~GeV. The strongest constraints are placed on heavy-neutral-lepton masses in the range 15--30~GeV of $|U_{e}|^2 < 1.1 \times 10^{-5}$ and $|U_{\mu}|^2 < 5 \times 10^{-6}$.
}

\AtlasRefCode{EXOT-2019-35}

\PreprintIdNumber{CERN-EP-2025-164}

\AtlasDate{\today}

\arXivId{2508.20929}

\HepDataRecord{167367}

\AtlasJournalRef{Eur.\ Phys.\ J.\ C\ 86\ (2026)\ 153}
\AtlasDOI{10.1140/epjc/s10052-025-15191-w}

\AtlasCoverEgroupAnalysisTeam{atlas-exot-2019-35-analysis-team@cern.ch}

\hypersetup{pdftitle={ATLAS document},pdfauthor={The ATLAS Collaboration}}

\begin{document}

\maketitle

\tableofcontents

\section{Introduction} %

%
\graphicspath{{figures/signal/}}

The discovery of neutrino flavour oscillation~\cite{Super-Kamiokande:1998kpq, SNO:2002tuh, KamLAND:2002uet} implies that neutrinos have  mass. This might indicate the existence of right-handed heavy Majorana neutrinos (HNL in the following, denoting a \enquote{heavy neutral lepton}) giving rise to the  so-called Type-1 Seesaw mechanism~\cite{Minkowski:1977sc,Yanagida:1979as,Glashow:1979nm,Gell-Mann:1979vob,Mohapatra:1979ia,Schechter:1980gr,Schechter:1981cv}. In this scenario the Standard Model (SM) neutrinos would acquire masses inversely proportional to the HNL mass, which would make them naturally small. The existence of HNLs can also be used to explain the baryon asymmetry of the universe via leptogenesis~\cite{Davidson:2008bu,Pilaftsis:2009pk,Shaposhnikov:2009zzb}. One of the HNL states can also be a viable candidate for dark matter~\cite{Asaka:2005an,Asaka:2005pn,Boyarsky:2009ix,Boyarsky:2018tvu,Ghiglieri:2020ulj} in models with more than one HNL.

HNLs may participate in weak interactions through mixing with the left-handed neutrinos via dimensionless mixing coefficients $U_{\alpha}$, with lepton flavours $\alpha\equiv{e, \mu, \tau}$. Realistic models able to explain neutrino oscillations would involve at least two HNLs with mixing to all SM flavours~\cite{Shaposhnikov:2006nn,Kersten:2007vk,Tastet:2021vwp}. It is, however, customary for experimental searches to use a simplified model with a single HNL mixing with a single lepton flavour, where the reach can be represented in the two-dimensional plane of the HNL mass
and a single active mixing parameter.  These searches can then be reinterpreted in different types of models~\cite{Tastet:2021vwp}.

For HNL masses up to 80~GeV, the main mechanism for HNL production in proton--proton ($pp$) collisions at the LHC is through mixing with the SM neutrinos produced in $W$ boson decays. From the experimental point of view, the most favourable HNL decay is the fully leptonic one ($N \rightarrow\ell^{\pm}\ell^{\mp}\nu$). The HNL decay can take place either promptly or after the HNL traverses a measurable distance in the detector (signifying a long-lived particle). Searches for this final state at the LHC have been published by the ATLAS, CMS and LHCb Collaborations, for both the prompt and long-lived particle signatures~\cite{ATLAS:2015gtp, ATLAS:2019kpx, LHCb:2020wxx, ATLAS:2022atq, CMS:2022fut, CMS:2023jqi, CMS:2024xdq}. Prior to the LHC, an analysis by the DELPHI Collaboration at LEP1 using \textasciitilde $10^6$ neutrinos from $Z$ boson decays provided the strongest direct constraints in the HNL mass range 2--75~GeV~\cite{delphi1997}.

The search documented in this paper targets the prompt decay of single HNLs produced in $W$ boson decays, using the full LHC \RunTwo data set (140~fb$^{-1}$) collected by the ATLAS experiment. It covers a wider HNL mass range than the ATLAS prompt decay search performed on a partial \RunTwo data sample~\cite{ATLAS:2019kpx}, and complements the results of the ATLAS search for long-lived HNLs~\cite{ATLAS:2022atq}. A CMS search for prompt HNL decays in their full \RunTwo data set was published recently in Ref.~\cite{CMS:2024xdq}.

\begin{figure}[th!]
\begin{center}
\begin{subfigure}{0.4\textwidth}
\includegraphics[width=1.\textwidth]{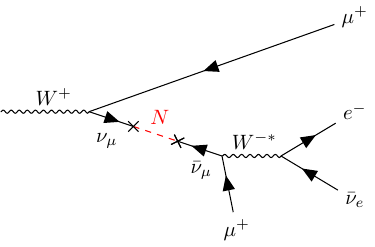}
\caption{ }
\label{fig:fig_01a}
\end{subfigure}\vspace{0.25cm}
\begin{subfigure}{0.4\textwidth}
\includegraphics[width=1.\textwidth]{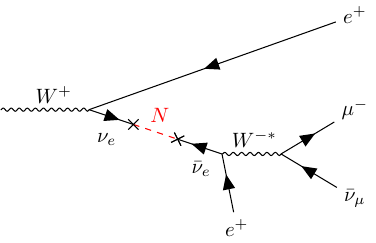}
\caption{ }
\label{fig:fig_01b}
\end{subfigure}
\caption{Feynman diagrams for heavy neutral lepton ($N$) production and decay in the channels originating from:
\protect\subref{fig:fig_01a}~HNL--$\nu_\mu$ mixing and $\mu^+ e^- \bar{\nu}_e$ decay of the heavy lepton, and \protect\subref{fig:fig_01b}~HNL--$\nu_e$ mixing and $e^+ \mu^- \bar{\nu}_\mu$ decay of the heavy lepton. %
Only lepton-number-violating processes are considered.
}
\label{fig:HNLfeynman}
\end{center}
\end{figure}

As illustrated in Figure~\ref{fig:HNLfeynman}, the mixing of HNLs with SM neutrinos allows them to be produced in $W$ boson decays, together with a charged SM lepton. The HNL then decays to a SM charged lepton and neutrino via charged EW currents. This process results in a final state comprising three charged leptons and a neutrino.
In this configuration, the cross section, branching fractions and kinematics of the process are determined solely by two parameters: a single active mixing parameter $U_{\ell}$ (with $\ell \equiv{e, \mu}$) and the mass of the heavy neutrino, $m_{\mathrm{HNL}}$.

The final state arising from the leptonic decay of the virtual $W$ boson is not constrained by the mixing of the HNL, so the three leptons can have the same flavour (SF) or two different flavours (DF). The possible charge/flavour combinations of the leptons arise from the fact that the HNL considered here is a Majorana particle. This analysis concentrates on final states with three leptons, two of which have the same flavour and the same electric charge (SC) so as to focus on lepton-number-violating processes. The third lepton is selected to have a different flavour and an opposite charge. Events with three SF leptons are not considered, because final states with opposite-charge (OC) SF leptons suffer from significant background contamination, originating from diboson $WZ$ production or Drell--Yan $\ell^+\ell^-$ production. These backgrounds contain e.g. a prompt third lepton (from $WZ$ processes where both bosons decay into leptons, or $ZZ$ and $VVV$ processes with leptonic $\tau$ decays) or a fake or non-prompt third lepton (i.e.\ a hadron faking a lepton, a lepton from a hadron decay, or an electron from a photon conversion that satisfies the prompt-lepton selection criteria presented in Section~\ref{sec:objects}). Such backgrounds are suppressed by requiring the two SF leptons to have the same charge.


%
%

%
\section{ATLAS detector}  %

%

\newcommand{\AtlasCoordFootnote}{%
ATLAS uses a right-handed coordinate system with its origin at the nominal interaction point (IP)
in the centre of the detector and the \(z\)-axis along the beam pipe.
The \(x\)-axis points from the IP to the centre of the LHC ring,
and the \(y\)-axis points upwards.
Polar coordinates \((r,\phi)\) are used in the transverse plane,
\(\phi\) being the azimuthal angle around the \(z\)-axis.
The pseudorapidity is defined in terms of the polar angle \(\theta\) as \(\eta = -\ln \tan(\theta/2)\) and is equal to the rapidity
\(y = (1/2) \ln[(E + p_z)/(E - p_z)]\) in the relativistic limit.
Angular distance is measured in units of \(\Delta R \equiv \sqrt{(\Delta y)^{2} + (\Delta\phi)^{2}}\).}

The ATLAS experiment~\cite{PERF-2007-01} at the LHC is a multipurpose particle detector
with a forward--backward symmetric cylindrical geometry and nearly \(4\pi\) coverage in
solid angle.\footnote{\AtlasCoordFootnote}
It consists of an inner detector surrounded by a thin superconducting solenoid
providing a \qty{2}{\tesla} axial magnetic field, electromagnetic and hadronic calorimeters, and a muon spectrometer.
The inner detector covers the pseudorapidity range \(|\eta| < 2.5\).
It consists of silicon pixel, silicon microstrip, and transition radiation tracking (TRT) detectors.
Lead/liquid-argon (LAr) sampling calorimeters provide electromagnetic (EM) energy measurements
with high granularity within the region \(|\eta|< 3.2\).
A steel/scintillator-tile hadronic calorimeter covers the central pseudorapidity range (\(|\eta| < 1.7\)).
The endcap and forward regions are instrumented with LAr calorimeters
for EM and hadronic energy measurements up to \(|\eta| = 4.9\).
The muon spectrometer surrounds the calorimeters and is based on
three large superconducting air-core toroidal magnets with eight coils each.
The field integral of the toroids ranges between \num{2.0} and \qty{6.0}{\tesla\metre}
across most of the spectrometer.
The muon spectrometer includes a system of precision tracking chambers up to \(|\eta| = 2.7\) and fast detectors for triggering up to \(|\eta| = 2.4\).
The luminosity is measured mainly by the LUCID-2~\cite{LUCID2} detector, which is located close to the beampipe.
A two-level trigger system is used to select events~\cite{TRIG-2016-01}.
The first-level trigger is implemented in hardware and uses a subset of the detector information
to accept events at a rate close to \qty{100}{\kHz}.
This is followed by a software-based trigger that
reduces the rate of accepting complete events to \qty{1.25}{\kHz} on average
depending on the data-taking conditions.
A software suite~\cite{SOFT-2022-02} is used in data simulation, in the reconstruction
and analysis of real and simulated data, in detector operations, and in the trigger and data acquisition
systems of the experiment.


%
%

%
%
%
%
%

%
\section{Data and MC simulations} %

\label{sec:samples}

The results presented in this paper are obtained using $pp$ collision data collected during \RunTwo of the LHC at a centre-of-mass energy of $\sqrt{s} = \SI{13}{\TeV}$. The number of $pp$ interactions per bunch-crossing (\pileup) in this data set ranges from about 8 to 70, with an average of 34~\cite{DAPR-2018-01}. Events recorded when parts of the detector were either not functional, or reserved for detector commissioning or calibration purposes are ignored, leaving 95.6\% of the recorded data~\cite{DAPR-2018-01} available for analysis. The integrated luminosity of this data set amounts to \SI{140}{fb^{-1}}, with an associated uncertainty of 0.83\%~\cite{DAPR-2021-01}, obtained using the LUCID-2 detector for the primary luminosity measurements, complemented by measurements using the inner detector and calorimeters.

Monte Carlo (MC) event simulations are mainly used to predict the background contributions from SM processes with prompt leptons, as well as those from hypothetical signal processes. Prompt leptons are defined as leptons that are not produced in the decay of a hadron, in the fragmentation of quarks and gluons, or in the conversion of a photon. MC samples are also used to validate the assumptions employed in data-based background estimation methods and to assess systematic uncertainties.

MC events were processed through a detailed simulation of the ATLAS detector~\cite{SOFT-2010-01}, based on \GEANT~\cite{Agostinelli:2002hh}. In some cases, a fast simulation~\cite{SOFT-2010-01} relying on a parameterization of the calorimeter response~\cite{ATL-PHYS-PUB-2010-013} was used instead. Additional minimum-bias interactions generated by \PYTHIA[8.186] using the \NNPDF[2.3lo] set of parton distribution functions (PDF)~\cite{Ball:2012cx} and the A3 set of tuned parameters~\cite{ATL-PHYS-PUB-2016-017} were simulated separately and overlaid on each simulated hard-interaction event to account for \pileup effects. The response of the detector and its electronic readout chain was then simulated~\cite{SOFT-2010-01}, also accounting for effects from interactions in the previous and following bunch-crossings. Reconstructed events are reweighted to reproduce the measured distributions of \pileup interactions in different data-taking periods. Reconstructed objects are further corrected for reconstruction inefficiencies.

%
%
%
%
%

\begin{table}[th!]
\caption{
A list of Monte Carlo event generators along with their settings for the primary simulated SM processes. When no reference for the cross-section normalization is provided, the value computed by the generator is used. LO and NLO refer to leading-order and next-to-leading-order calculations in the strong coupling $\alpha_s$, respectively. In certain cases (as indicated), the accuracy of the matrix element depends on the number of additional parton emissions.
}
\label{tab:generators}\vspace{-0.25cm}
\scriptsize
\renewcommand{\arraystretch}{1.2}
\resizebox{\textwidth}{!}{
\begin{tabular}{l|l|l|l|l|l|l}
\hline\hline
Process & Generator & Computation & Parton shower & Cross-section & PDF set & Set of tuned \\
&           & order       &               & normalization &         & parameters  \\
\hline
\hline
Diboson~\cite{ATL-PHYS-PUB-2017-005}
& \SHERPA[2.2.2]~\cite{Bothmann:2019yzt}
& NLO 0-1j + LO 2-3j
& \textsc{CSShower}~\cite{Gleisberg:2008fv,Schumann:2007mg}
& NLO
& \NNPDF[3.0nnlo]~\cite{Ball:2014uwa}
& default \\
& + \OPENLOOPS~\cite{Buccioni:2019sur,Cascioli:2011va,Denner:2016kdg}
& & & & & \\
Triboson~\cite{ATL-PHYS-PUB-2017-005}
& \SHERPA[2.2.1]~\cite{Bothmann:2019yzt}
& LO 0-1j
& \textsc{CSShower}~\cite{Gleisberg:2008fv,Schumann:2007mg}
& NLO
& \NNPDF[3.0nnlo]~\cite{Ball:2014uwa}
& default \\
$t\bar{t}W$~\cite{ATL-PHYS-PUB-2020-024}
& \SHERPA[2.2.10]~\cite{Bothmann:2019yzt}
& NLO 0-1j + LO 2j
& \textsc{CSShower}~\cite{Schumann:2007mg}
& NLO
& \NNPDF[3.0nnlo]~\cite{Ball:2014uwa}
& default \\
& + \OPENLOOPS~\cite{Buccioni:2019sur,Cascioli:2011va,Denner:2016kdg}
& + LO $\mathcal{O}(\alpha^3\alphas$)
& & & & \\
$t\bar{t}\ell^\pm \ell^\mp$, $m_{\ell\ell}>\SI{5}{GeV}$~\cite{ATL-PHYS-PUB-2020-024}
& \SHERPA[2.2.1]~\cite{Bothmann:2019yzt}
& NLO
& \textsc{CSShower}~\cite{Gleisberg:2008fv,Schumann:2007mg}
& NLO
& \NNPDF[3.0nnlo]~\cite{Ball:2014uwa}
& default \\
\quad\quad\quad $1<m_{\ell\ell}<\SI{5}{GeV}$
& MG5\_\AMCatNLO[2.3.3]~\cite{Alwall:2014hca}
& NLO
& \PYTHIA[8.212]~\cite{Sjostrand:2007gs}
& NLO
& \NNPDF[3.0nlo]~\cite{Ball:2014uwa}
& A14~\cite{ATL-PHYS-PUB-2014-021} \\
$t\bar{t}H$~\cite{ATL-PHYS-PUB-2016-005}
& \POWHEGBOX[v2]~\cite{Bothmann:2019yzt}
& NLO
& \PYTHIA[8.230]~\cite{Sjostrand:2007gs}
& NLO~\cite{deFlorian:2016spz}
& \NNPDF[3.0nlo]~\cite{Ball:2014uwa}
& A14~\cite{ATL-PHYS-PUB-2014-021} \\
\hline
$\ttbar$~\cite{ATL-PHYS-PUB-2016-020}
& \POWHEGBOX[v2]~\cite{Frixione:2007nw,Nason:2004rx,Frixione:2007vw,Alioli:2010xd}
& NLO
& \PYTHIA[8.230]~\cite{Sjostrand:2014zea}
& NNLO~\cite{ATL-PHYS-PUB-2016-020}
& \NNPDF[3.0nlo]~\cite{Ball:2014uwa}
& A14~\cite{ATL-PHYS-PUB-2014-021} \\
Single top ($s$-, $t$-channel)
& \POWHEGBOX[v2]~\cite{Re:2010bp,Nason:2004rx,Frixione:2007vw,Alioli:2010xd}
& NLO
& \PYTHIA[8.230]~\cite{Sjostrand:2014zea}
& NNLO~\cite{Campbell:2020fhf,PDF4LHCWorkingGroup:2022cjn}
& \NNPDF[3.0nlo]~\cite{Ball:2014uwa}
& A14~\cite{ATL-PHYS-PUB-2014-021} \\
\phantom{single top }($tW$)
& \SHERPA[2.2.7]~\cite{Bothmann:2019yzt}
& NLO
& \textsc{CSShower}~\cite{Gleisberg:2008fv,Schumann:2007mg}
& NNLO + NNLL~\cite{Kidonakis:2021vob}
& \NNPDF[3.0nnlo]~\cite{Ball:2014uwa}
& default \\
$W\to\ell^\pm \nu$, $Z/\gamma^*\to\ell^\pm \ell^\mp$~\cite{ATL-PHYS-PUB-2017-006}
& \SHERPA[2.2.11]~\cite{Bothmann:2019yzt}
& NLO 0-2j + LO 3-5j
& \textsc{CSShower}~\cite{Gleisberg:2008fv,Schumann:2007mg}
& NNLO~\cite{Anastasiou:2003ds,ATL-PHYS-PUB-2017-006}
& \NNPDF[3.0nnlo]~\cite{Ball:2014uwa}
& default \\
\hline\hline
\end{tabular}
}
\end{table}


Table~\ref{tab:generators} presents the MC event generators and the corresponding settings used to generate the SM process samples. This includes the selected parton shower algorithms, the tuned parameter sets, and the PDF sets. When using \PYTHIA, the decays of bottom and charm hadrons were simulated using the \EVTGEN program~\cite{Lange:2001uf}. Diboson ($VV$) processes~\cite{ATL-PHYS-PUB-2017-005} encompass all resonant and non-resonant $pp\to 3\ell\nu/4\ell/\ell^{\pm}\ell^{\pm}\nu\nu$ processes of order $\alpha^4$ in the fine-structure constant, including contributions from the Higgs boson, as well as vector-boson scattering/fusion processes at order $\alpha^6$. Triboson ($VVV$) processes include all relevant resonant and non-resonant processes with up to six charged leptons in the final state at order $\alpha^6$. The process $pp \to \ttbar Z$, with the $Z$ boson decaying into a pair of same-flavour opposite-charge (SFOC) leptons, was generated for dilepton invariant masses as low as \SI{1}{\GeV}. Samples of simulated $\ttbar$ and single-top events were also produced. The diagram removal scheme~\cite{Frixione:2008yi} was used to account for overlaps between the \ttbar and $tW$ samples.

Other processes not specifically listed in the table but considered for background estimates include the $4t$, $tH$, $tWH$, $\ttbar ZZ$, $\ttbar WH$, and $\ttbar HH$ processes. All these samples were produced using a fast detector simulation.

The HNL signal MC samples were generated with \MADGRAPH[2.9.3]~\cite{Alwall:2014hca} and the \NNPDF[2.3lo] PDF, using the \AtlasMC{HeavyN} model~\cite{Alva:2014gxa,Degrande:2016aje} at LO in QCD, which also allows the emission of up to two additional partons. \PYTHIA[8.245] with the A14 set of tuned parameters~\cite{ATL-PHYS-PUB-2014-021} was used to model the parton showering, hadronization, and underlying event. Matrix element to parton shower matching was performed using the CKKW-L prescription~\cite{Lonnblad:2011xx}. The $W$ boson was set to decay exclusively into a muon or electron, and an HNL: $W \rightarrow \mu^{\pm} N$ or $W \rightarrow e^\pm N$. Only $N \to \mu^{\pm} e^{\mp} \bar{\nu}_{e}$ or $N\to e^\pm \mu^\mp \bar{\nu}_{\mu}$ decays were simulated. Thus, $W$ and $N$ decays to $\tau$-leptons or jets are not included.\footnote{Using MC simulations it is found that, for $\tau$-leptons produced in HNL decays, the contribution from leptonic decays of the $\tau$-lepton after the lepton selection process is less than 5\%. A much smaller contribution is expected after the full signal region requirements are applied.} Separate samples were generated for the two neutrino flavour mixing parameters, $U_{e}$ and $U_{\mu}$, as illustrated in Figure~\ref{fig:HNLfeynman}.

Signal samples were produced individually for HNL masses of 8, 10, 15, 20, 30, 40, 50, 60, and 65~GeV, with the mean proper decay length set to $c\tau = 0.1$~mm for all samples. For the lowest HNL mass samples of 8, 10, and 15~GeV, a decay length of $c\tau = 1$~mm was also simulated. Since the low-mass signal samples were generated only for two HNL decay lengths, the signal reconstruction efficiency is computed for other decay lengths by a reweighting of the lifetime distributions of these samples, following the methodology from Ref.~\cite{EXOT-2017-25}. Each $c\tau = 0.1$~mm sample is reweighted to multiple increments of decay length: from 0.1 to 0.5~mm in steps of 0.05~mm, and from 0.5 to 1.0~mm in steps of 0.1~mm. A similar reweighting scheme is applied to the $c\tau = 1$~mm samples, which are used to validate the intermediate lifetime reconstruction from the $c\tau = 0.1$~mm samples; additional details are in Section~\ref{sec:uncertainties}.

A fast detector simulation is used for all signal samples. The product of the cross section for $W$ boson production in $\sqrt{s} =13$~TeV $pp$ collisions and the branching ratio for leptonic $W$ boson decay into a single lepton flavour~\cite{Atre:2009rg,Gorbunov:2007ak} is taken from the ATLAS measurement in Ref.~\cite{ATAS:V_xsec} as $\sigma(pp \rightarrow W) \times B(W \rightarrow \ell \nu) = 20.6 \pm 0.7$~nb. The total decay width of the HNL and the corresponding lifetime are computed as detailed in Refs.~\cite{Atre:2009rg,Gorbunov:2007ak}.


%
%

%
\section{Object identification and event reconstruction}  %

\label{sec:objects}

Charged-particle tracks within $\abseta<2.5$ are reconstructed~\cite{ATLAS-CONF-2012-042,Salzburger:2015sgq,ATL-PHYS-PUB-2018-002} in the ATLAS inner detector and subsequently combined to create primary vertex candidates that are constructed using at least two tracks~\cite{PERF-2015-01,ATL-PHYS-PUB-2015-026}. Among these, the primary vertex is identified as the vertex with the largest $\sum\left(\pt^\text{track}\right)^2$, where $\pt^\text{track}$ is the transverse momentum of a track associated with the vertex. The transverse and longitudinal impact parameters of all tracks, denoted by $d_0$ and $z_0$, are calculated relative to the primary vertex~\cite{IDTR-2019-05}.
Jets with $\abseta<4.5$ are reconstructed using the \textsc{FastJet} implementation~\cite{Fastjet} of the \antikt algorithm~\cite{Cacciari:2008gp}, with a radius parameter of $R=0.4$. The inputs to this algorithm are particle-flow objects~\cite{JETM-2018-05,PERF-2015-09}, which combine measurements from the inner detector and calorimeters~\cite{PERF-2014-07} to enhance the jet energy resolution and increase the jet reconstruction efficiency, particularly at low jet transverse momentum. Calibrations are applied to the jet mass, transverse momentum (\pt), energy scale and energy resolution which include components derived both from simulation and in situ measurements, documented thoroughly in Ref.~\cite{JETM-2018-05}. Only jets with $\pt>\SI{20}{\GeV}$ and $\abseta<2.8$ are retained. Events containing reconstructed jets induced by calorimeter noise or non-collision backgrounds, identified using criteria similar to those described in Ref.~\cite{ATLAS-CONF-2015-029}, are removed. Jets originating from pile-up interactions, according to a track-based discriminant (JVT)~\cite{PERF-2014-03}, are rejected.

Within the inner-detector acceptance, jets containing bottom hadrons (\bjets) are identified using the DL1r tagging algorithm~\cite{FTAG-2019-07}, which uses the properties of reconstructed tracks and secondary vertices. The analysis selects true \bjets with an estimated efficiency of 85\%, as measured in a $\ttbar$-enriched event sample~\cite{FTAG-2019-07}. The DL1r algorithm is calibrated using a likelihood-based method for each jet type~\cite{FTAG-2018-01}, and correction factors are applied to the simulated event samples to account for differences between data and simulation in the \bjet tagging efficiencies for $b$-jets, $c$-jets, and light-flavour jets.

Electron candidates are reconstructed from clustered energy deposits in the electromagnetic calorimeter matched to an inner-detector track re-fitted to account for bremsstrahlung losses~\cite{EGAM-2018-01,EGAM-2021-01}. The electron momentum is determined by a calibration procedure based on boosted decision trees (BDTs)~\cite{EGAM-2021-02}. Only the electrons satisfying the requirements $|\eta|<2.47$ and $\pt>\SI{8}{\GeV}$ are used, and electrons within the transition region $1.37<|\eta|<1.52$ between the barrel and endcap calorimeters are discarded. Electrons from background sources are rejected with a likelihood discriminant~\cite{EGAM-2018-01,EGAM-2021-01} built from information about the development of the electron shower in the calorimeter, its compatibility with the matched track, and particle identification in the TRT detector. The electron candidates must satisfy the \enquote{Loose} identification criterion described in Ref.~\cite{EGAM-2021-01}. They must also fulfil a requirement on the transverse impact parameter divided by its uncertainty: $|d_{0}|/\sigma(d_{0}) < 5$. The electron track $z_0$ is required to satisfy $|z_{0}\sin\theta| < \SI{0.5}{mm}$, where $\theta$ is the polar angle of the track. Electron candidates satisfying these requirements are referred to as baseline electrons.

Signal electrons are defined as baseline electrons that satisfy the \enquote{Medium} identification criterion~\cite{EGAM-2021-01}. This tighter identification criterion is imposed to further suppress fake electrons arising from misidentified jets, as well as non-prompt electrons from decays of hadrons.
The identification requirements are complemented by isolation criteria that reject electrons with significant energy in a cone around the electron candidate, calculated using either non-electron tracks or energy clusters. The efficiency of the applied \enquote{Loose\_VarRad} isolation criterion rises with increasing \pt, from 76\% at approximately $\pt = \SI{10}{GeV}$ to 99\% at $\pt = \SI{100}{GeV}$, as measured using $Z \to ee$ events~\cite{EGAM-2021-01}. Signal electrons must not be associated with the vertex of a reconstructed photon conversion in the detector material~\cite{EGAM-2018-01,EGAM-2021-01}. To further reduce the photon conversion background in selections with two SC leptons, additional requirements are applied to the signal electrons~\cite{Aad:2020klt,EGAM-2018-01}: the electron candidate must not have a displaced vertex reconstructed at a radius $r > \SI{20}{mm}$, where the reconstruction uses the track associated with the electron; and the invariant mass of the system formed by the electron track and the closest track at the primary vertex or a conversion vertex is required to exceed $\SI{100}{MeV}$. This combined selection is referred to as the photon conversion veto. Electrons that are very likely to have a wrongly assigned charge are identified and subsequently rejected using the ECIDS discriminant~\cite{EGAM-2018-01}, a BDT based on the properties of the electron track, accepting 98\% of simulated $Z\to ee$ decay electrons while rejecting 90\% of those with the wrong charge.

Muon candidates are obtained from an iterative track fit applied to inner detector and muon spectrometer hits~\cite{MUON-2018-03}. Momentum corrections are applied to compensate for detector misalignments~\cite{MUON-2022-01}. Only candidates with $|\eta|<2.5$ and $\pt>\SI{8}{\GeV}$ are considered, and they must satisfy the \enquote{Medium} quality criteria defined in Ref.~\cite{MUON-2018-03}.
Muons must fulfil $|\zzsth|<\SI{0.5}{mm}$, to reject muon candidates from \pileup. Candidates satisfying these requirements are referred to as baseline muons. Approximately 0.1\% of events contain a muon with poorly estimated momentum, and such events are rejected. Signal muons are defined as baseline muons that also satisfy $|d_{0}|/\sigma(d_{0}) < 3$, along with the \enquote{TightTrackOnly\_VarRad} isolation criterion detailed in Ref.~\cite{MUON-2018-03}, to further suppress fake/non-prompt muons.
To avoid interpreting the same detector signals as multiple objects, an overlap removal procedure is applied to baseline leptons and jets. Jets within $\Delta{R}=0.2$ of an electron or muon are removed. Leptons that are closer than $\Delta{R}=\operatorname{min}(0.4,0.1+\SI{9.6}{\GeV}/\pt(\ell))$ to any remaining jet are discarded.
The missing transverse momentum, \ptmiss, and its magnitude, \met, are reconstructed~\cite{ATLAS-CONF-2018-023,JETM-2020-03} from lepton candidates, jets, and reconstructed photons ($\pt>\SI{25}{\GeV}$, $|\eta|<2.37$) that meet the \enquote{Tight} identification requirements~\cite{EGAM-2021-01}. In addition, a track-based \enquote{soft term} composed of inner detector tracks linked to the primary vertex but excluded from the previously mentioned objects is included. The \met reconstruction employs its own overlap removal procedure~\cite{ATLAS-CONF-2018-023,JETM-2020-03}.

Several variables are defined in order to maximize the sensitivity to the HNL signal:
\begin{itemize}
\item The transverse momentum of the leading lepton (i.e.\ the highest-$\pt$ signal lepton), denoted by \ptLeadL.
\item The minimum invariant mass of a different-flavour opposite-charge (DFOC) signal lepton pair, $\mDFOSmin$, in events with a DFOC lepton pair.
\item The invariant mass of the SC signal lepton pair, $\mLLSS$.
\item The invariant mass of the three signal leptons, \mlll.
\item The size of the interval of masses of the three-lepton and $\ptmiss$ system that results in no real solutions for the reconstructed event kinematics, $\mWCompMax - \mWCompMin$; see Section~\ref{sec:selection} for the details of the computation.
\item The transverse mass, $\mt = \sqrt{2 \ptLeadL \met(1 - \cos(\Delta\phi(\ell1,\ptmiss)))}$, computed using the $\met$ and the leading signal lepton in the event.
\item The inclusive effective mass, \meff, defined as the scalar sum of the \pt of all jets and leptons, as well as the \met.
\item The distance in $\eta$--$\phi$ between the two SC signal leptons, $\DRLLSS$.
\item The azimuthal distance $\DPhillMet$ between the dilepton system (defined by the leading and sub-leading signal leptons) and the missing transverse momentum.
\item The number of signal leptons in the event, \NSigLep.
\item The number of baseline leptons in the event, \NBaseLep.
\item The number of \bjets in the event, \NBjetsTF.
\end{itemize}


%
%

%
\section{Event selection}  %

\label{sec:selection}

A combination of dilepton triggers is used to select the events~\cite{TRIG-2018-05,ATLAS:2020gty}, and the offline lepton candidates are matched to the trigger objects. Offline electrons activating the $ee$ trigger must have \pt above 18~GeV, while offline muons activating the $\mu\mu$ trigger must have \pt above 9~GeV and 19~GeV in the case of the asymmetric dilepton trigger, and above 15~GeV for the symmetric trigger. For the $e \mu$ asymmetric triggers, electrons and muons  must have \pt above 8~GeV and 25~GeV (respectively) or 18~GeV and 15~GeV (respectively).

An event preselection is defined by the following criteria. The leading and sub-leading signal leptons must have $\pt$ of at least 20 and 15~GeV, respectively. Events must have at most three baseline leptons, to suppress backgrounds from $\ttbar Z$ and $\ttbar H$ events. Moreover, at least one same-flavour same-charge (SFSC) signal lepton pair must be present in the event, to further suppress background processes such as $ZZ$ and $VVV$. Events with three SF signal leptons are rejected. To suppress $W/Z$+jets, $VV$, $\mDFOSmin$ is required to be $>5$~GeV and $\mLLSS$ must be $>20$~GeV. The scalar sum of \pt for all signal leptons must be $>50$~GeV, to reduce backgrounds with fake/non-prompt leptons. To reduce $W/Z$+jets, $WZ$ and $\ttbar$ backgrounds, \meff must be $<250$~GeV. Finally, requiring $\met<100$~GeV suppresses mainly backgrounds containing a $W\to\ell\nu$ decay (e.g.\ \ttbar\ events).

Signal regions (SRs) are defined by taking each signal sample separately and optimising the expected significance using a range of kinematic selections. The significance is calculated using background expectations computed from MC events, as well as using the full background estimates. In cases where the optimal requirements were close for similar $m_{\mathrm{HNL}}$, the corresponding SR definitions were merged to end up with less regions. The SRs can overlap, and in the final stages of the analysis (see Section~\ref{sec:results}), only the SR with the highest expected sensitivity for a given HNL mass is used.

%
\begin{table}[th!]
\caption{
Signal regions considered in the analysis.
}
\label{tab:Final_SRs}\vspace{-0.25cm}
\scriptsize
\def\arraystretch{1.2}
\begin{center}
\resizebox{.95\textwidth}{!}{
\begin{tabular}{l|c|c|c|c|c|c|c|c|l}
\hline \hline
SR   & \NSigLep & $\mt$ & $\meff$ & $\DRLLSS$ & $\mDFOSmin$ & $\mLLSS$ & $\mWCompMax - \mWCompMin$ & \NBjetsTF & Additional   \\
name &          & [GeV] &  [GeV]  &           & [GeV]       & [GeV]    & [GeV]                     &           & requirements \\
\hline\hline
\multicolumn{10}{l}{Three-lepton SRs} \\ %
\hline
SRE1   & $=3$ & $<70$ & $<250$ & $>2.0$ & $<20$ & $<75$ & $<60$ & $=0$ & - \\
SRE2   & $=3$ & $<65$ & $<200$ & $-$    & $<25$ & $<65$ & $<60$ & $=0$ & - \\
SRE3   & $=3$ & $<50$ & $<150$ & $>1.5$ & $<35$ & $<60$ & $<70$ & $=0$ & - \\
SRE4   & $=3$ & -     & $<250$ & $-$    & $<50$ & -     & $<70$ & $=0$ & $\mlll < 80$ GeV, $\ptLeadL < 30$~GeV  \\
\hline
SRM1   & $=3$ & $<70$ & $<120$ & $>2.0$ & $<25$ & $<70$ & $<70$ & $=0$ & - \\
SRM2   & $=3$ & $<45$ & $<120$ & $-$    & $<40$ & $<65$ & $<70$ & $=0$ & $\DPhillMet>1.0$ \\
\hline
\multicolumn{10}{l}{Two-lepton SRs} \\ %
\hline
SRM3   & $=2$ (SC) & $<60$ & $<120$ & $>2.5$ & $-$ & $<75$ & - & -  & -  \\
SRM4   & $=2$ (SC) & $<55$ & $<150$ & $>2.0$ & $-$ & $<70$ & - & $=0$ & -  \\
SRM5   & $=2$ (SC) & $<70$ & $<120$ & $>2.0$ & $-$ & $<75$ & - & -  & -  \\
\hline\hline
\end{tabular}}
\end{center}
\end{table}
%


The SRs are presented in Table~\ref{tab:Final_SRs}. They are defined separately for $e^\pm e^\pm \mu^\mp$ (labelled SRE$i$, $i\equiv{1 \to 4}$) and $\mu^\pm \mu^\pm e^\mp$ (labelled SRM$j$, $j\equiv{1 \to 2}$) final states with three signal leptons, and $\mu^\pm \mu^\pm e$ (labelled SRM$j$, $j\equiv{3 \to 5}$) final states with two SFSC signal leptons and one additional baseline electron.
The main sensitivity of the analysis comes from the signal regions with three signal leptons ($\NSigLep=3$), while the signal two-lepton signal regions (\NSigLep=2) recover some events where an electron passes the baseline but fails the signal requirements.
In the three-lepton SRs, events with three leptons of the same electric charge are removed. In the two-lepton SRs a requirement on the charge of the third baseline lepton is found to have no effect on the SM background, and therefore is not applied.
Because some SRs overlap, for a given $m_{\mathrm{HNL}}$ only one of the three-lepton SRE$i$ ($i\equiv{1 \to 4}$) is picked to search for the $e^\pm e^\pm \mu^\mp$ signature, while one three-lepton SRM$j$ ($j \equiv{1 \to 2}$) and one two-lepton SRM$j$ ($j \equiv{3 \to 5}$) are used together for the $\mu^\pm \mu^\pm e^\mp$ signature.

The requirements on the $\DRLLSS$ variable in the SR definition suppress the $W/Z$+jets background.
For HNL signal events, the invariant mass of the three leptons and the neutrino should be compatible with the decay of a $W$ boson. When the transverse momentum of the neutrino is identified with the \ptmiss\ vector, the missing $z$-component can be computed by imposing a mass constraint $\mWComp$ for the three leptons and neutrino system. The resulting quadratic equation has real solutions only if the discriminant, which is itself a quadratic function of $\mWComp$, is $\geq 0$. In case the discriminant is smaller than zero for $\mWComp = 80.4$~GeV, the two values are calculated ($\mWCompMax$ and $\mWCompMin$) for which the discriminant is zero. In case the discriminant is positive, $\mWCompMax$ and $\mWCompMin$ are taken to be zero. Large values of $\mWCompMax - \mWCompMin$  are observed for backgrounds such as $t\bar{t}$ and $VV$, whereas for the HNL signal, much lower values are seen. This is because, for the backgrounds, the invariant mass of the system comprising the three leptons and \ptmiss\ is typically very far from the $W$-boson mass.
Finally, the requirement on the \NBjetsTF variable in the SR definition in Table~\ref{tab:Final_SRs} helps to reduce backgrounds from sources that have at least one $b$-jet in the decay chain, such as \ttbar events.

To validate the background estimates in the SRs, the requirements on some of the kinematic variables used to define the SRs are relaxed; the events falling into any SR are subsequently removed from the VRs ensuring no overlap. These regions, referred to as validation regions (VRs), are defined in Table~\ref{tab:Relaxed_SRs}, and have a background composition similar to that in the signal regions. The level of agreement between the observed data and the estimated background in these validation regions is shown in Figure~\ref{fig:Bkg_Val_VRs} and discussed in Section~\ref{sec:background_val}.

%
\begin{table}[th!]
\caption{
Definitions of the VRs used to validate the background expectations. Events in the VRs satisfy the preselection requirements, as well as the requirements on \NSigLep, \NBaseLep, and \NBjetsTF applied for the corresponding SRs, and differ from the SRs by the requirements shown here.
}
\label{tab:Relaxed_SRs}\vspace{-0.25cm}
\scriptsize
\def\arraystretch{1.2}
\begin{center}
\resizebox{0.85\textwidth}{!}{
\begin{tabular}{l|c|c|c|c|c|c|l}
\hline \hline
VRs & $\mt$ & $\meff$ & $\DRLLSS$ & $\mDFOSmin$ & $\mLLSS$ & $\mWCompMax - \mWCompMin$ & Additional    \\
& [GeV] &  [GeV]  &           & [GeV]       & [GeV]    & [GeV]                     & requirements \\
\hline\hline
All VRs & \multicolumn{7}{l}{Veto events belonging to any SR} \\
\hline
\multicolumn{8}{l}{Three-lepton VRs} \\ %
\hline
VRE1 & $<80$ & $<250$ & $>2.0$ & $<50$ & $<75$ & $<80$ & - \\
VRE2 & $<80$ & $<200$ & -      & $<50$ & $<75$ & $<80$ & - \\
VRE3 & $<80$ & $<150$ & $>1.5$ & $<50$ & $<75$ & $<80$ & - \\
VRE4 & -     & $<250$ & -      &  -    & -     & $<70$ & \ptLeadL $<55$~GeV, \mlll cut removed \\
\hline
VRM1 & [70, 80] & $<250$ & $>2.0$             & $<50$ & $<75$ & $<80$ & - \\
VRM2 & [45, 80] & $<150$ & -                  & $<50$ & $<75$ & $<80$ & $\DPhillMet>1.0$ \\
\hline
\multicolumn{8}{l}{Two-lepton VRs} \\ %
\hline
VRM3 & $<70$ & $<250$ & $>1.0$ & - & $<75$ & - & - \\
VRM4 & $<70$ & $<250$ & $>1.0$ & - & $<70$ & - & - \\
VRM5 & $<70$ & $<250$ & $>1.5$ & - & $<75$ & - & - \\
\hline\hline
\end{tabular}}
\end{center}
\end{table}



%
%

%
\section{Background estimation}   %

\label{sec:backgrounds}
\graphicspath{{figures/background/}}

The background contributions to the final states in this analysis can be categorized into three groups: SM processes that produce genuine prompt leptons in the final state; SM processes that result in same-charge lepton pairs due to the misidentification of the charge of one of the electrons; and SM processes that result in SC pairs or three leptons due to fake/non-prompt leptons. The estimation methods for these categories are outlined in the following three subsections. MC simulations are used for processes with three prompt leptons, while the other two categories rely on data events selected with specific lepton criteria.
\subsection{SM processes with three prompt leptons}

Due to the \bjet veto present in most of the signal regions, as well as the low \met and \meff requirements applied during the event preselection, processes that include top quarks are highly suppressed. As a result, the largest prompt-lepton backgrounds in the SRs come from the $WZ$ process, with both bosons decaying into leptons (including leptonically decaying $\tau$-leptons). The $VVV$ and $ZZ/WW$ processes also contribute to the prompt-lepton background category in their direct decay to three light leptons (e.g., $W^+~W^+~W^- \rightarrow \mu^+ \nu_{\mu} ~\mu^+ \nu_{\mu} ~e^- \bar{\nu}_{e}$), as well as when they decay to $\tau$-leptons, which subsequently decay to light leptons. To avoid double-counting the sources of fake/non-prompt-lepton background, events with fake/non-prompt leptons are removed from MC simulations by using generator-level lepton information. The contributions from these SM background processes are estimated by normalizing the MC samples to their theoretical cross sections.

\subsection{Electrons with misidentified charge}
\label{sec:background_flip}

The charge of an electron is given by the sign of the curvature of its track in the ATLAS inner detector. However, if the electron radiates a photon that converts and multiple tracks are reconstructed in the inner detector, an incorrect charge may be attributed (\enquote{charge-flip}). The probability $\xi$ for signal electrons to undergo a charge-flip ranges between $0.03\%$ and $0.07\%$, and varies as a function of $|\eta|$ and \pt, as illustrated in Ref.~\cite{EGAM-2018-01}.

In this analysis, the charge-flip background is found to be negligible in all signal regions. However, charge-flip contributes to some control regions used in Section~\ref{sec:background_fnp} and in certain two-signal-lepton selections used for the background validation presented in Section~\ref{sec:background_val}. The contributions to the $e^\pm e^\pm$ and $e^\pm \mu^\pm$ final states are estimated by selecting data events containing OC leptons and weighting them according to the known $\xi(|\eta|,\pt)$ values. These values are derived from simulated $Z \to ee$ events and are adjusted by correction factors $\gamma(|\eta|,\pt) = \xi_\mathrm{data}/\xi_\mathrm{MC}$ to correct for known mismodelling. The correction factors $\gamma$ and their associated uncertainties, assumed to be process-independent, are obtained~\cite{EGAM-2018-01} from comparisons of OC and SC dielectron pair rates observed in $Z\to ee$ decays in data and MC events. These factors are found to lie within 20\% of unity, regardless of the electron $\eta$ and \pt.

The dominant uncertainties in the predicted charge-flip yields arise from the measurement of the $\gamma$ corrections, which is statistically limited and also influenced by significant background contamination~\cite{EGAM-2018-01}. The predicted yields have a typical uncertainty of 40\%.

\subsection{Fake and non-prompt leptons}
\label{sec:background_fnp}

Fake/non-prompt leptons~\cite{EGAM-2019-01} are defined as either a hadron faking a lepton, a lepton from a hadron decay, or an electron from a photon conversion.
When combining this fake/non-prompt-lepton candidate with one or two prompt leptons in an event, a same-charge lepton-pair or three-lepton signature may be formed. According to MC simulation, fake-lepton or non-prompt-lepton background in the SRs defined with two SC signal leptons comes primarily from $W/Z$+jets processes, with a small contribution from \ttbar events. In the SRs defined with three signal leptons, the main background sources are $Z$+jets and \ttbar processes.
Backgrounds from these sources are estimated using the data-based matrix method.

The matrix method~\cite{D0:1999qdf,EGAM-2019-01,SUSY-2020-27} leverages the differing efficiencies of identification and isolation criteria when applied to fake/non-prompt leptons instead of prompt leptons. Within a specific region of interest (such as the SRs), data events are selected using lepton selection criteria that are looser than those defining the signal leptons described in Section~\ref{sec:objects}. These events are subsequently categorized according to the number of signal leptons they contain. A fully determined system of linear equations can then be constructed~\cite{EGAM-2019-01,SUSY-2020-27}, relating the counts of such categorized events to the unknown numbers of events containing only prompt leptons, exactly one fake/non-prompt lepton, and so forth. The coefficients in these equations are functions of the probabilities $\varepsilon(|\eta|,\pt)$ and $\zeta(|\eta|,\pt)$, representing the probabilities that loose prompt leptons or fake/non-prompt leptons, respectively, also satisfy the signal lepton criteria.

The sample of loosely selected leptons comprises the subset of baseline leptons after overlap removal that also satisfy a $|d_0|/\sigma(d_0) <7$ requirement for muons, and the ECIDS criterion for electrons along with all selections aimed at removing the electrons due to photon conversions, as detailed in Section~\ref{sec:objects}. The estimated contribution of charge-flip electrons described in Section~\ref{sec:background_flip} is subtracted as detailed in Refs.~\cite{EGAM-2019-01,SUSY-2018-09,SUSY-2020-27}.

The probabilities $\varepsilon(|\eta|,\pt)$ are calculated with simulated $\ttbar$ events containing semileptonic top-quark decays, as recommended in Ref.~\cite{EGAM-2019-01}. These MC-based measurements are corrected~\cite{EGAM-2019-01} for known mismodelling by applying the representative scale factors given in Refs.~\cite{EGAM-2021-01,MUON-2022-01}. For both muons and electrons, $\varepsilon$ is found to increase with \pt, from around 75\% at \SI{10}{GeV} to 99\% at \SI{200}{GeV}. A dependency on $\eta$ is accounted for, although it is observed to be weak.
For electrons, systematic uncertainties are as large as $5.5\%$ at low $\pt$ and decrease to below $1\%$ in the $\pt > 25$~GeV region.
For muons, systematic uncertainties vary between $1\%$ and $2\%$ in the $\pt < 25$~GeV range, decreasing to less than $0.6\%$ in higher \pt ranges. For both electrons and muons, the systematic uncertainties in $\varepsilon$ are dominated by the uncertainties associated with the scale factors in Refs.~\cite{EGAM-2021-01,MUON-2022-01}.

To measure the probabilities $\zeta(|\eta|,\pt)$, the approach presented in Refs.~\cite{EGAM-2019-01,SUSY-2018-09,SUSY-2020-27} is used. Data control regions enriched in fake/non-prompt leptons are defined in order to select events containing one or two prompt leptons and a fake or non-prompt lepton that together form a same-charge pair. In such regions, signal contamination is reduced to a negligible amount by applying the requirement $\mDFOSmin > 30$~GeV. The measurement is done within the range $8<\pt<\SI{75}{GeV}$ for electrons and $8<\pt<\SI{40}{GeV}$ for muons. Separate measurements are made for events with two SC signal leptons and events with three signal leptons, to account for the leading contributors to the fake/non-prompt-lepton background in the SRs.

For electrons, $\zeta(|\eta|,\pt)$ is found to increase with \pt, varying between 5\% and 30\%.
For muons, $\zeta(|\eta|,\pt)$ is found to decrease with \pt, varying between 6\% and 20\% in the most relevant range of $8<\pt<\SI{40}{GeV}$.
In the SRM3, SRM4 and SRM5 signal regions, which select events containing a third baseline lepton in addition to the pair of same-charge signal leptons, the fake/non-prompt-lepton background estimation makes the assumption~\cite{SUSY-2020-27} that the fake/non-prompt lepton is part of the SC pair.

The same conservative systematic uncertainties as in Refs.~\cite{SUSY-2018-09,SUSY-2020-27} are used to account for contamination from prompt same-charge leptons in the measurement regions and for the assumption that $\varepsilon$ and $\zeta$ can be used outside of the regions in which they are measured.
This procedure leads to uncertainties in $\zeta$ ranging from $20\%$ at lower \pt to 50\% for high-\pt leptons of both flavours. These systematic uncertainties combined with statistical uncertainties in the $\zeta$ measurements~\cite{EGAM-2019-01} are taken as systematic uncertainties in the predicted fake/non-prompt-lepton yields, and are between 28\% and 43\% in the signal regions.

\subsection{Validation of the background estimates}
\label{sec:background_val}

\begin{figure}[th!]
\begin{center}
\includegraphics[width=\textwidth]{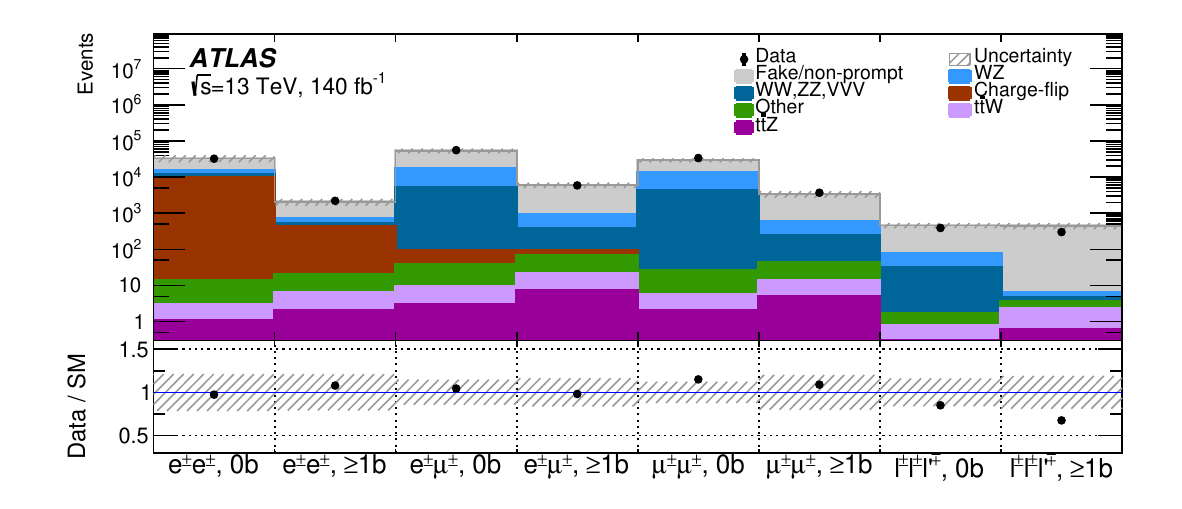}
\end{center}
\caption{
Comparison between the data and estimated background at preselection level. Events entering the SRs defined in Section~\ref{sec:selection} are vetoed.
The events are classified in terms of the number of leptons and their flavours, as well as the number of \bjets.
The $\ell^\pm \ell^\pm$ bins have a $\geq 2$ signal leptons selection, with no requirement on the number of baseline leptons; the $\ell^\pm \ell^\pm \ell'^\mp$ bins have a $=3$ signal leptons selection.
The uncertainties shown with hashed bands include only the statistical uncertainties and the full uncertainties associated with the data-driven background estimates.
The bottom panel shows the ratio of the observed data yields to the predicted background yields.
}
\label{fig:Bkg_Val_Presel}
\end{figure}

To check the validity and robustness of the background estimates, observed data are compared with the predicted background after the event preselection and excluding events in the signal regions presented in Section~\ref{sec:selection}. Figure~\ref{fig:Bkg_Val_Presel} shows this comparison across different lepton-flavour and \bjet-multiplicity combinations, with generally good agreement confirming the validity of the matrix method in estimating the fake/non-prompt-lepton background as well as the methods used to estimate the electron charge-flip and prompt SM backgrounds. The signal contamination is negligible.

\begin{figure}[ht!]
\begin{center}
\includegraphics[width=\textwidth]{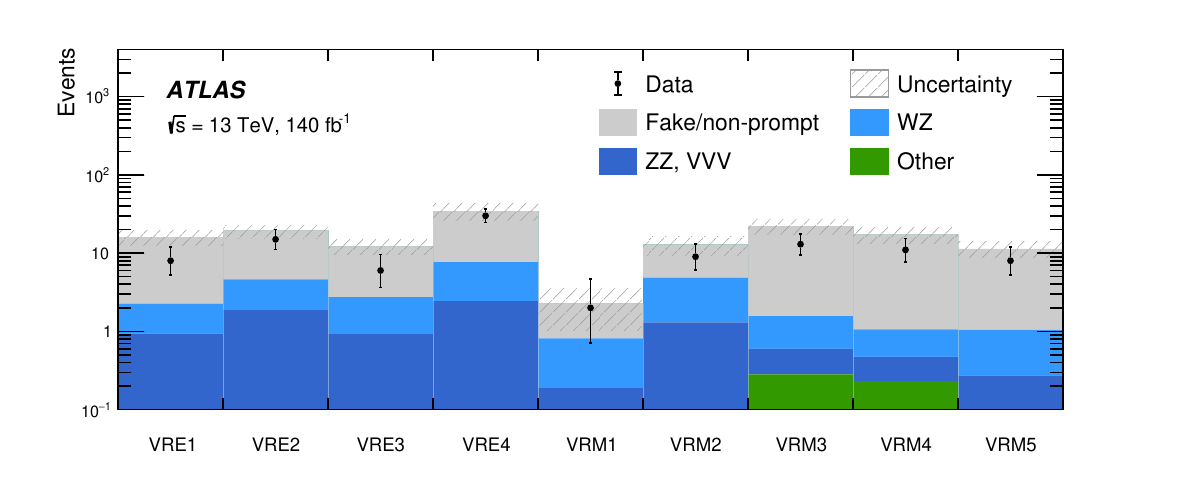}
\end{center}
\caption{
Comparison between the data and estimated background in the validation regions. The hatched band represents the total uncertainty in the estimated background.
}
\label{fig:Bkg_Val_VRs}
\end{figure}

Figure~\ref{fig:Bkg_Val_VRs} shows the results in the validation regions, defined in Table~\ref{tab:Relaxed_SRs} to be closer to the signal regions by applying tighter requirements on the kinematic variables than at preselection level. The observed data and expected background agree in all VRs within 2 sigma, both when considering all sources of uncertainties, and only the statistical uncertainties. A signal at the previous exclusion bounds presented in Ref.~\cite{ATLAS:2019kpx} would lead to contamination in the validation regions between 4\% and 13\%, with the largest values reached in VRM2.


%
%

%
\section{Systematic uncertainties} %

\label{sec:uncertainties}

Several sources of systematic uncertainty are accounted for in the analysis.
The uncertainty in the integrated luminosity of the combined \RunTwo data set is $0.83\%$~\cite{DAPR-2021-01},
and affects the normalization of all simulated samples.
Differences between the pile-up distributions in MC simulation and data are minimized by means of MC event reweighting, and a corresponding uncertainty is computed by increasing and decreasing the mean number of simultaneous interactions by 4\%.
The uncertainties related to the overall electron selection efficiency arise from the electron energy scale and resolution, and the trigger, reconstruction, identification and isolation efficiencies, and are obtained from Refs.~\cite{EGAM-2021-01,TRIG-2018-05}. The muon-related uncertainties arise from the muon momentum scale and resolution, and the trigger, reconstruction, identification and isolation efficiencies, as well as the track-to-vertex matching, and are obtained from Refs.~\cite{MUON-2018-03,TRIG-2012-03}.
Uncertainties in the jet energy scale (JES) are derived by combining information from test-beam data,
LHC collision data and MC simulation~\cite{JETM-2018-05}. Uncertainties in the jet energy resolution (JER) are
estimated as a function of the jet transverse momentum and rapidity using dijet events~\cite{JETM-2018-05}.
Uncertainties related to the jet vertex tagger (JVT)~\cite{ATL-PHYS-PUB-2019-026} and
flavour tagging~\cite{FTAG-2018-01,FTAG-2020-08,FTAG-2019-02} are applied.
Uncertainties in the \met value are estimated by propagating the uncertainties in the energy or momentum scale for each of the objects entering the calculation, as well as the uncertainties in the soft-term resolution and scale~\cite{ATLAS-CONF-2018-023}.
All the above uncertainties are treated as fully correlated among the analysis signal regions and the physics processes considered.

The systematic uncertainty in the signal production cross section is estimated to be 3.7\%, dominated by the uncertainty in the measured $W$ boson production cross section~\cite{ATAS:V_xsec}.
Systematic uncertainties arising from the signal's quadratic dependence on the mixing angle and the phase-space factors in the HNL decay are negligible, and therefore not considered in this analysis. Acceptance uncertainties arising from the renormalization scale (\muR), factorization scale (\muF) and PDF scale choices are evaluated by either halving or doubling the values of these scales. PDF uncertainties, including the effect of \alphas uncertainty, are assessed by following the PDF4LHC15 prescription~\cite{Butterworth:2015oua}. A uniform 18\% uncertainty is applied to the low-mass HNL signal points to take account of uncertainties in the lifetime reweighting technique discussed in Section~\ref{sec:results}. This is evaluated by comparing the efficiency predicted by the reweighted $c\tau = 0.1$~mm sample with that from the reweighted $c\tau = 1.0$~mm sample.

The uncertainties associated with backgrounds from sources such as electrons with misidentified charge and fake or non-prompt leptons are discussed in Section~\ref{sec:backgrounds}.
For background contributions from $VV$/$VVV$ production, an overall uncertainty of 60\% are estimated using the method described in Ref.~\cite{STDM-2018-03}.
For $t\bar{t}W$, $t\bar{t}Z$ and $t\bar{t}H$ production, an overall uncertainty of 50\% is assigned to each process to account for cross-section and modelling uncertainties~\cite{SUSY-2020-27}. An overall uncertainty of 50\% is assigned to the rare processes $t\bar{t}HH$, $t\bar{t}VV$, $tV$, $tVV$, $t\bar{t}t$ and $t\bar{t}t\bar{t}$.

In all signal regions, the dominant sources of uncertainty are the statistical and systematic uncertainties in the estimation of the fake/non-prompt-lepton background. Subdominant sources include the theoretical systematic uncertainties of the $VV$ and $VVV$ backgrounds. The statistical uncertainties in the MC background yields range from 19\% to 30\% in the SRE$i$ signal regions and from 12\% to 22\% in the SRM$j$ signal regions.


%
%

%
\section{Results}  %

\label{sec:results}
\graphicspath{{figures/TREX_yields/}{figures/TREX_SRs/}{figures/TREX_limits/}}

The event yields in the various signal regions are presented in Section~\ref{TREX:yields_SRs}. In Section~\ref{TREX:Discovery_limits}, discovery fits are performed in order to quantify any signal-like excess above the background prediction. With no significant excesses observed, the results are interpreted in Section~\ref{TREX:MD_limits} as limits on the mixing parameters.

\subsection{Event yields in the signal regions}
\label{TREX:yields_SRs}

The observed number of events in each SR along with the background predictions and uncertainties are shown in Table~\ref{table:prefit_SRs_results_summary}. The contribution from the fake/non-prompt-lepton background dominates in all the SRs.
As mentioned in Section~\ref{sec:selection}, the signal regions partially overlap in the following groups: SRE$i$, $i \equiv{1 \to 4}$; SRM$j$, $j \equiv{1 \to 2}$; and SRM$j$, $j \equiv{3 \to 5}$.
Therefore, the same data events can be found in more than one signal region, e.g.\ seven data events are selected in both SRE1 and SRE2, and five data events are selected in both SRE1 and SRE3. The largest excess of observed events over the background prediction, 1.7 standard deviations, appears in SRE3. When counting the events present in all the SRs (without double-counting), a total of 44 data events and $30 \pm 5$ predicted background events are obtained, corresponding to an excess of 1.7 standard deviations.
%

\begin{table}[ht!]
\caption{
The observed number of data events and the expected background contributions in the SRs.
The uncertainties are both statistical and systematic.
The individual uncertainties can be correlated or anticorrelated and therefore do not necessarily add in quadrature to equal the total uncertainty.
}
\label{table:prefit_SRs_results_summary}\vspace{-0.25cm}
\scriptsize
\def\arraystretch{1.2}
\centering
\resizebox{1\textwidth}{!}{
\begin{tabular}{l|cccc|cc|ccc}
\hline\hline
& SRE1 & SRE2 & SRE3 & SRE4 & SRM1 & SRM2 & SRM3 & SRM4 & SRM5 \\
\hline\hline
Observed & 8 & 10 & 15 & 10 & 6 & 9 & 5 & 7 & 10 \\
Total background & \numRF{4.37114}$\pm$ \numRF{1.52457} & \numRF{5.09081}$\pm$ \numRF{1.48397} & \numRF{7.85031}$\pm$ \numRF{2.11933} & \numRF{5.95799}$\pm$ \numRF{1.71751} & \numRF{3.72657}$\pm$ \numRF{1.18977} & \numRF{6.22073}$\pm$ \numRF{1.85462} & \numRF{5.08798}$\pm$ \numRF{1.56874} & \numRF{6.15551}$\pm$ \numRF{1.66467} & \numRF{7.26419}$\pm$ \numRF{1.8499} \\
\hline
$WZ$ & \numRF{0.197614}$\pm$ \numRF{0.133671} & \numRF{0.252284}$\pm$ \numRF{0.178352} & \numRF{0.59145}$\pm$ \numRF{0.396829} & \numRF{0.6}$\pm$ \numRF{0.4} & \numRF{0.697619}$\pm$ \numRF{0.431385} & \numRF{0.71129}$\pm$ \numRF{0.452746} & \numRF{0.27689}$\pm$ \numRF{0.176741} & \numRF{0.240267}$\pm$ \numRF{0.160925} & \numRF{0.520664}$\pm$ \numRF{0.329559} \\
$ZZ$, $VVV$ & \numRF{0.394375}$\pm$ \numRF{0.249038} & \numRF{0.880699}$\pm$ \numRF{0.593969} & \numRF{0.758141}$\pm$ \numRF{0.518029} & \numRF{0.370477}$\pm$ \numRF{0.248858} & \numRF{0.677375}$\pm$ \numRF{0.416937} & \numRF{1.10423} $\pm$ \numRF{0.699151} & \numRF{0.162918}$\pm$ \numRF{0.104707} & \numRF{0.230911}$\pm$ \numRF{0.153022} & \numRF{0.295892}$\pm$ \numRF{0.185714} \\
Others & $<0.05$ & $<0.05$ & $<0.05$ & $<0.05$ & $<0.05$ & $<0.05$ & $<0.05$ & $<0.05$ & $<0.05$ \\
Fake/non-prompt & \numRF{3.753}$\pm$ \numRF{1.49578} & \numRF{3.931}$\pm$ \numRF{1.33092} & \numRF{6.472}$\pm$ \numRF{1.99556} & \numRF{5.004}$\pm$ \numRF{1.64248} & \numRF{2.346}$\pm$ \numRF{1.01931} & \numRF{4.399}$\pm$ \numRF{1.63739} & \numRF{4.646}$\pm$ \numRF{1.55391} & \numRF{5.679}$\pm$ \numRF{1.64783} & \numRF{6.444}$\pm$ \numRF{1.80781} \\
\hline \hline
\end{tabular}}
\end{table}


%
\subsection{Discovery fit results}
\label{TREX:Discovery_limits}

A discovery fit is performed separately for each signal region defined in Section~\ref{sec:selection} using a profile-likelihood-ratio test, as detailed in Ref.~\cite{Baak:2014wma}. The likelihood function $L(\mu, \vec{\theta})$ is the Poisson probability for the observed number of events given the sum of the predicted signal and background yields, $S+B$. The value of $S$ depends on the signal strength ($\mu$). In addition, both $S$ and $B$ depend on a set of nuisance parameters ($\vec{\theta}$) which include both the systematic and statistical uncertainties of the predicted signal and background yields~\cite{Cranmer:2012sba}.
Each nuisance parameter has an associated constraint term. For systematic uncertainties, this constraint term is a Gaussian. Statistical uncertainties, resulting from the limited size of the simulated samples, are assessed using the Beeston--Barlow \enquote{lite} technique~\cite{Conway:2011in}, which employs a Poisson constraint. Systematic uncertainties are treated as correlated. The fitting procedure maximizes the likelihood by varying the signal strength and nuisance parameters to extract their best-fit values.

The test statistic $\lambda = -2\ln(L(\mu, \hat{\hat{\vec{\theta}}})/L(\hat{\mu}, \hat{\vec{\theta}}))$ is evaluated with the RooFit package~\cite{RooFit,Baak:2014wma,TREX_2025}. Here $\hat{\mu}$ and $\hat{\vec{\theta}}$ are the parameter values  that maximize the likelihood function, and $\hat{\hat{\vec{\theta}}}$ are the parameter values that maximize the likelihood function for a fixed value of $\mu$. The discovery fit for each SR is used to assess whether the observed data event yield is incompatible with the background-only hypothesis. For this, $\lambda$ is calculated with $\mu=0$ while the fit allows $\hat{\mu}$ to vary to account for the possible presence of a signal in the data. A corresponding distribution of $\lambda$ is generated using the predicted event model under the background-only hypothesis with the asymptotic formulae given in Ref.~\cite{Cowan:2010js}. From this distribution a $p$-value is calculated using the $\lambda$ value derived from the observed data. This $p$-value reflects the probability that the observed data event yield is compatible with the background-only hypothesis.
%

%
%
%
%

\begin{table}[ht!]
\caption{
The discovery fit results. For SRM3, where the expected background slightly exceeds the observed number of data events (see Table~\ref{table:prefit_SRs_results_summary}), the $p$-value is capped at 0.5, corresponding to a significance of zero.
}
\label{tab:discoveryAllSRs}\vspace{-0.25cm}
\renewcommand{\arraystretch}{1.2}
\centering
\resizebox{0.85\textwidth}{!}{
\begin{tabular}{l|cccc|cc|ccc}
\hline\hline
& SRE1 & SRE2 & SRE3 & SRE4 & SRM1 & SRM2 & SRM3 & SRM4 & SRM5 \\
\hline\hline
$p$-value              & \numRF{0.1133500} & \numRF{0.0611279} & \numRF{0.0413460} & \numRF{0.1150785} & \numRF{0.1848968} & \numRF{0.2067143} & %
0.5	& \numRF{0.3805851} & \numRF{0.1924470} \\
Significance [$\sigma$]	 & \numRF{1.2089030} & \numRF{1.5453738} & \numRF{1.7352751} & \numRF{1.1999540} & \numRF{0.8968600} & \numRF{0.8178748} & $0$             & \numRF{0.3039443} & \numRF{0.8689140} \\
\hline\hline
\end{tabular}}
\end{table}


The discovery fit results are shown in Table~\ref{tab:discoveryAllSRs}. Most signal regions show moderate tension between the observed data and the background prediction. The difference is most notable in the SRE$i$ signal regions.
\subsection{Model-dependent limits}
\label{TREX:MD_limits}

In the absence of a significant excess over the SM expectation, model-dependent exclusion limits are derived using the CL$_\mathrm{s}$
method~\cite{Junk:1999kv,Read:2002hq} with the statistical tools described in Section~\ref{TREX:Discovery_limits}.
Limits are placed on the mixing parameters $|U_{e}|^2$ and $|U_{\mu}|^2$ for the HNL mass range 8--65~GeV. For a given signal
scenario, values of the mixing-parameter strengths ($\mu$) yielding $\mathrm{CL_s}< 0.05$, where CL$_\mathrm{s}$ is computed using the
asymptotic approximation~\cite{Cowan:2010js}, are excluded at $\geq$ 95\% CL. For the observed limit, the CL$_\mathrm{s}$ is computed using the test statistic calculated with the observed data set, while for the expected limit the CL$_\mathrm{s}$ is computed using the test statistic calculated from an Asimov data set with nuisance parameters set to those extracted from a background-only fit to data~\cite{Baak:2014wma}. For each signal mass point, the signal region that minimizes the expected limit is used to quote the observed and expected upper limits on $\mu$. For the $\mu^\pm \mu^\pm e^\mp$ signal channel, the most sensitive two-lepton and three-lepton signal regions are combined.

\begin{figure}[th!]
\begin{center}
\begin{subfigure}{0.45\textwidth}
\includegraphics[width=1.\textwidth]{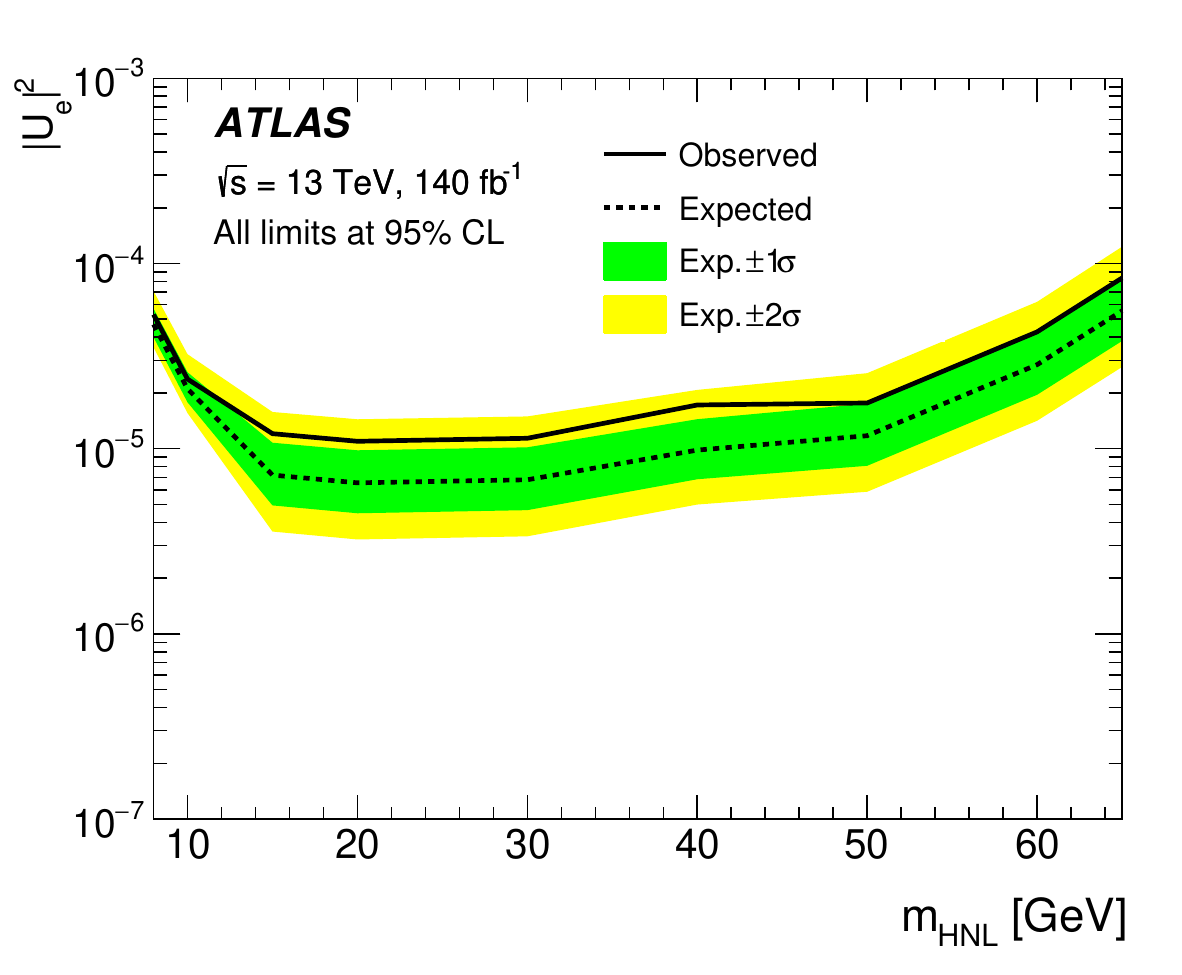}
\caption{ }
\label{fig:SR_HNL_08_65_limit_eemu_EEM}
\end{subfigure}
\begin{subfigure}{0.45\textwidth}
\includegraphics[width=1.\textwidth]{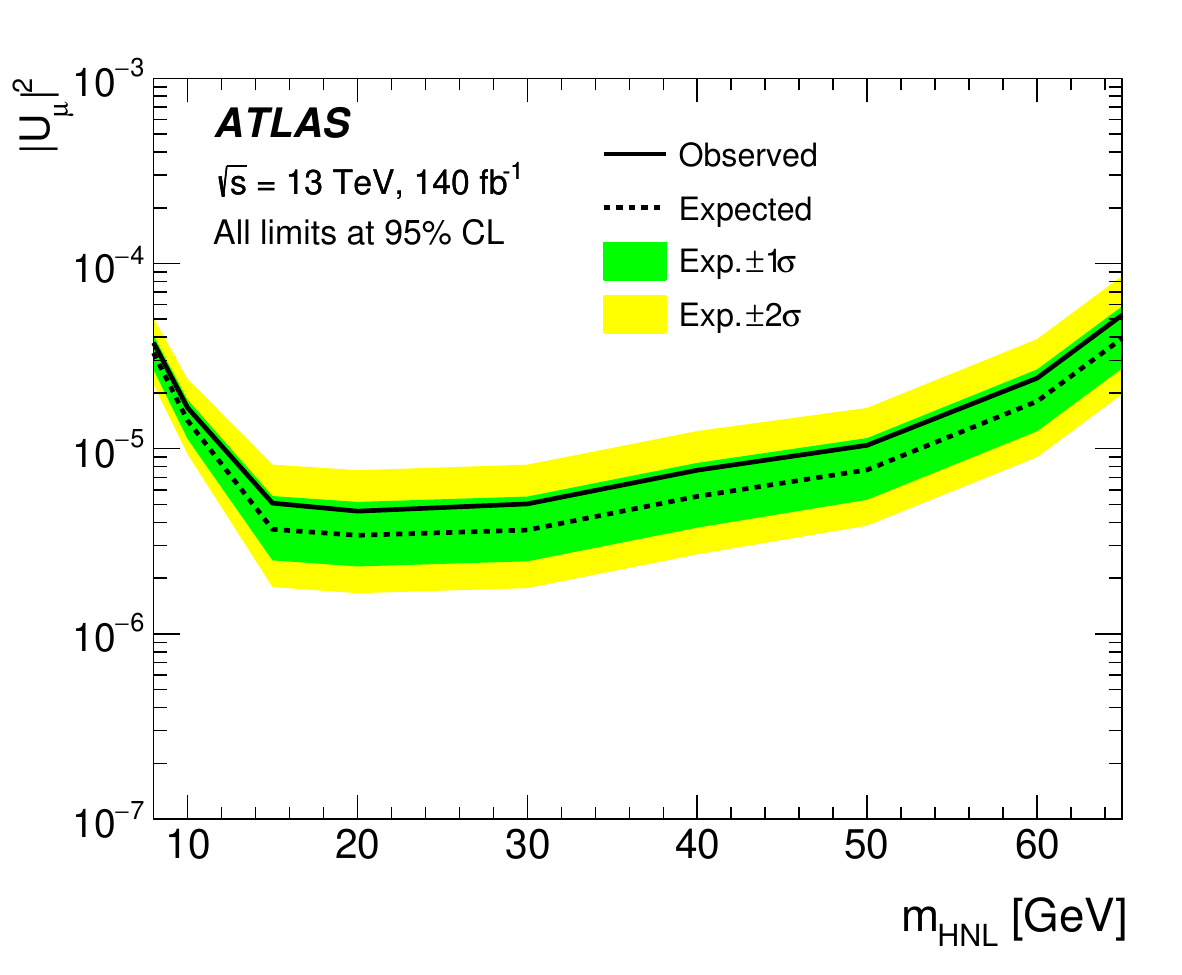}
\caption{ }
\label{fig:SR_HNL_08_65_limit_mumue_MME}
\end{subfigure}
\end{center}
\caption{
Observed 95\% confidence level (CL) exclusion limits for the \protect\subref{fig:SR_HNL_08_65_limit_eemu_EEM}~$|U_{e}|^2$ and \protect\subref{fig:SR_HNL_08_65_limit_mumue_MME}~$|U_{\mu}|^2$ mixing parameters versus the HNL mass.
The expected (dashed line) exclusion limits are also shown.
The $1\sigma$ and $2\sigma$ uncertainty bands around the expected exclusion limit reflect uncertainties in signal and background yields.
}
\label{fig:MD_Limits}
\end{figure}

The limits placed on the $|U_{e}|^2$ and $|U_{\mu}|^2$ mixing parameters as a function of the HNL mass are presented in Figure~\ref{fig:MD_Limits}.
For the $e^\pm e^\pm \mu^\mp$ signal channel, the SRE1 signal region is the most sensitive one for HNL masses $\leq10$~GeV. The SRE2 signal region is then used up to 30~GeV, while the SRE3 signal region is used for the 40~GeV HNL mass point. Finally, SRE4 is used for the results in the HNL high mass region ($\geq 50$~GeV). For the $\mu^\pm \mu^\pm e^\mp$ signal channel, SRM1 is the most sensitive three-lepton signal region in the HNL low mass range ($\leq 40$~GeV), while the SRM2 signal region is used in the HNL high-mass region.
The SRM3 and SRM4 two-lepton signal regions are used in the HNL low mass range: SRM3 for the $\leq 10$~GeV and 20~GeV mass points, and SRM4 for the 15, 30, and 40~GeV mass points.
SRM5 is used for HNL masses $\geq 50$~GeV.

The analysis excludes $|U_{e}|^2$ values above $1.1 \times 10^{-5}$ in the mass range 15 to 30~GeV, and $|U_{\mu}|^2$ values above $5.0 \times 10^{-6}$ in the range 20 to 30~GeV. Upper limits of $1.8 \times 10^{-5}$ and $7.0 \times 10^{-6}$ are set on $|U_{e}|^2$ and $|U_{\mu}|^2$ across wider mass ranges of 15 to 50~GeV and 15 to 40~GeV, respectively.

At higher masses, the sensitivity of the analysis decreases due to kinematic suppression of HNL production in $W$ boson decay.
The lifetime of the HNL is proportional to $m_N^{-5} \times |U_{\ell}|^{-2}$ and therefore increases at low HNL masses, resulting in efficiency loss due to the impact-parameter criteria  (see Section~\ref{sec:objects}).
As discussed in Section~\ref{sec:samples}, this efficiency loss is evaluated by reweighting the lowest HNL mass signal samples generated with $c\tau = 0.1$~mm to any intermediate value of $c\tau$. For each nominal and intermediate HNL decay length, a limit on the mixing parameter is derived. These limits are then interpolated using a cubic spline for each low mass point, which is subsequently used to extract the final limit on the mixing parameter.
As a result of the increasing rate of efficiency loss at lower HNL masses, the number of predicted signal events falls more rapidly with decreasing mixing parameters, leading to narrower uncertainty bands in the 8--15~GeV mass range.

The results of this analysis supersede those of the previous ATLAS search based on a partial \RunTwo data sample~\cite{ATLAS:2019kpx}.
They also complement the results of the ATLAS analysis addressing long-lived HNLs~\cite{ATLAS:2022atq}, by including the region from 20 to 65~GeV in the covered HNL mass range, as well as the results of the ATLAS displaced vertex search at lower masses~\cite{EXOT-2022-12}, and the ATLAS two leptons plus two or more jets~\cite{EXOT-2012-24} and vector-boson fusion two same-sign leptons~\cite{EXOT-2020-06, EXOT-2023-16} analyses at higher masses.


%
%

%
\section{Conclusion}    %

%

A search for heavy neutral leptons (HNLs) produced in leptonic decays of on-shell $W$ bosons is performed using data recorded
by the ATLAS detector at the LHC in $pp$ collisions at a centre-of-mass energy of $\sqrt{s} = \SI{13}{\TeV}$, corresponding to an integrated luminosity of 140~fb$^{-1}$. The search focuses on the HNL decay into two charged leptons and a neutrino, resulting in a final state with three prompt charged leptons (either muons or electrons). Exploiting the Majorana nature of the HNL to suppress SM background, the final state is required to have a same-flavour, same-charge lepton pair. To increase the efficiency, events in which the third lepton has lower quality (referred to as two-signal-lepton events) are also accepted.

The observed event yields are consistent with the background expectations. The results are presented as limits on the HNL mixing parameter as a function of HNL mass between 8 and 65~GeV. Two cases are considered, with the HNL mixing with either an electron neutrino (with mixing parameter $|U_{e}|^2$) or a muon neutrino (with mixing parameter $|U_{\mu}|^2$).
The analysis excludes $|U_{e}|^2$ values above $8\times 10^{-5}$ and $|U_{\mu}|^2$ values above $5.0 \times 10^{-5}$ in the full mass range of 8--65~GeV. The strongest constraints are placed on heavy-neutral-lepton masses in the range 15--30~GeV of $|U_{e}|^2 < 1.1 \times 10^{-5}$ and $|U_{\mu}|^2 < 5 \times 10^{-6}$.
%


\FloatBarrier
\section*{Acknowledgements}
%

%

%
%

%
%

We thank CERN for the very successful operation of the LHC and its injectors, as well as the support staff at
CERN and at our institutions worldwide without whom ATLAS could not be operated efficiently.

The crucial computing support from all WLCG partners is acknowledged gratefully, in particular from CERN, the ATLAS Tier-1 facilities at TRIUMF/SFU (Canada), NDGF (Denmark, Norway, Sweden), CC-IN2P3 (France), KIT/GridKA (Germany), INFN-CNAF (Italy), NL-T1 (Netherlands), PIC (Spain), RAL (UK) and BNL (USA), the Tier-2 facilities worldwide and large non-WLCG resource providers. Major contributors of computing resources are listed in Ref.~\cite{ATL-SOFT-PUB-2025-001}.

We gratefully acknowledge the support of ANPCyT, Argentina; YerPhI, Armenia; ARC, Australia; BMWFW and FWF, Austria; ANAS, Azerbaijan; CNPq and FAPESP, Brazil; NSERC, NRC and CFI, Canada; CERN; ANID, Chile; CAS, MOST and NSFC, China; Minciencias, Colombia; MEYS CR, Czech Republic; DNRF and DNSRC, Denmark; IN2P3-CNRS and CEA-DRF/IRFU, France; SRNSFG, Georgia; BMFTR, HGF and MPG, Germany; GSRI, Greece; RGC and Hong Kong SAR, China; ICHEP and Academy of Sciences and Humanities, Israel; INFN, Italy; MEXT and JSPS, Japan; CNRST, Morocco; NWO, Netherlands; RCN, Norway; MNiSW, Poland; FCT, Portugal; MNE/IFA, Romania; MSTDI, Serbia; MSSR, Slovakia; ARIS and MVZI, Slovenia; DSI/NRF, South Africa; MICIU/AEI, Spain; SRC and Wallenberg Foundation, Sweden; SERI, SNSF and Cantons of Bern and Geneva, Switzerland; NSTC, Taipei; TENMAK, T\"urkiye; STFC/UKRI, United Kingdom; DOE and NSF, United States of America.

Individual groups and members have received support from BCKDF, CANARIE, CRC and DRAC, Canada; CERN-CZ, FORTE and PRIMUS, Czech Republic; COST, ERC, ERDF, Horizon 2020, ICSC-NextGenerationEU and Marie Sk{\l}odowska-Curie Actions, European Union; Investissements d'Avenir Labex, Investissements d'Avenir Idex and ANR, France; DFG and AvH Foundation, Germany; Herakleitos, Thales and Aristeia programmes co-financed by EU-ESF and the Greek NSRF, Greece; BSF-NSF and MINERVA, Israel; NCN and NAWA, Poland; La Caixa Banking Foundation, CERCA Programme Generalitat de Catalunya and PROMETEO and GenT Programmes Generalitat Valenciana, Spain; G\"{o}ran Gustafssons Stiftelse, Sweden; The Royal Society and Leverhulme Trust, United Kingdom.

In addition, individual members wish to acknowledge support from CERN: European Organization for Nuclear Research (CERN DOCT); Chile: Agencia Nacional de Investigaci\'on y Desarrollo (FONDECYT 1230812, FONDECYT 1240864, Fondecyt 3240661, Fondecyt Regular 1240721); China: Chinese Ministry of Science and Technology (MOST-2023YFA1605700, MOST-2023YFA1609300), National Natural Science Foundation of China (NSFC - 12175119, NSFC 12275265); Czech Republic: Czech Science Foundation (GACR - 24-11373S), Ministry of Education Youth and Sports (ERC-CZ-LL2327, FORTE CZ.02.01.01/00/22\_008/0004632), PRIMUS Research Programme (PRIMUS/21/SCI/017); EU: H2020 European Research Council (ERC - 101002463); European Union: European Research Council (BARD No. 101116429, ERC - 948254, ERC 101089007), European Regional Development Fund (SMASH COFUND 101081355, SLO ERDF), Horizon 2020 Framework Programme (MUCCA - CHIST-ERA-19-XAI-00), European Union, Future Artificial Intelligence Research (FAIR-NextGenerationEU PE00000013), Italian Center for High Performance Computing, Big Data and Quantum Computing (ICSC, NextGenerationEU); France: Agence Nationale de la Recherche (ANR-21-CE31-0022, ANR-22-EDIR-0002, ANR-24-CE31-0504-01); Germany: Baden-Württemberg Stiftung (BW Stiftung-Postdoc Eliteprogramme), Deutsche Forschungsgemeinschaft (DFG - 469666862, DFG - CR 312/5-2); China: Research Grants Council (GRF); Italy: Istituto Nazionale di Fisica Nucleare (ICSC, NextGenerationEU), Ministero dell'Università e della Ricerca (NextGenEU 153D23001490006 M4C2.1.1, NextGenEU I53D23000820006 M4C2.1.1, NextGenEU I53D23001490006 M4C2.1.1, SOE2024\_0000023); Japan: Japan Society for the Promotion of Science (JSPS KAKENHI JP22H01227, JSPS KAKENHI JP22H04944, JSPS KAKENHI JP22KK0227, JSPS KAKENHI JP24K23939, JSPS KAKENHI JP25H00650, JSPS KAKENHI JP25H01291, JSPS KAKENHI JP25K01023); Norway: Research Council of Norway (RCN-314472); Poland: Ministry of Science and Higher Education (IDUB AGH, POB8, D4 no 9722), Polish National Science Centre (NCN 2021/42/E/ST2/00350, NCN OPUS 2023/51/B/ST2/02507, NCN OPUS nr 2022/47/B/ST2/03059, NCN UMO-2019/34/E/ST2/00393, UMO-2022/47/O/ST2/00148, UMO-2023/49/B/ST2/04085, UMO-2023/51/B/ST2/00920, UMO-2024/53/N/ST2/00869); Portugal: Foundation for Science and Technology (FCT); Spain: Ministry of Science and Innovation (MCIN \& NextGenEU PCI2022-135018-2, MICIN \& FEDER PID2021-125273NB, RYC2019-028510-I, RYC2020-030254-I, RYC2021-031273-I, RYC2022-038164-I), Ministerio de Ciencia, Innovación y Universidades/Agencia Estatal de Investigaci\'on (PID2022-142604OB-C22); Sweden: Carl Trygger Foundation (Carl Trygger Foundation CTS 22:2312), Swedish Research Council (Swedish Research Council 2023-04654, VR 2021-03651, VR 2022-03845, VR 2022-04683, VR 2023-03403, VR 2024-05451), Knut and Alice Wallenberg Foundation (KAW 2018.0458, KAW 2022.0358, KAW 2023.0366); Switzerland: Swiss National Science Foundation (SNSF - PCEFP2\_194658); United Kingdom: Leverhulme Trust (Leverhulme Trust RPG-2020-004), Royal Society (NIF-R1-231091); United States of America: U.S. Department of Energy (ECA DE-AC02-76SF00515), Neubauer Family Foundation.

%
%


%
%
%
%
%
%
%
%

%
%
%
\printbibliography
\clearpage
\input{atlas_authlist}
%

%
%
%
%

%
%
%
%

\end{document}

%% file: atlas_authlist.tex
 
\begin{flushleft}
\hypersetup{urlcolor=black}
{\Large The ATLAS Collaboration}

\bigskip

\AtlasOrcid[0000-0002-6665-4934]{G.~Aad}$^\textrm{\scriptsize 104}$,
\AtlasOrcid[0000-0001-7616-1554]{E.~Aakvaag}$^\textrm{\scriptsize 17}$,
\AtlasOrcid[0000-0002-5888-2734]{B.~Abbott}$^\textrm{\scriptsize 123}$,
\AtlasOrcid[0000-0002-0287-5869]{S.~Abdelhameed}$^\textrm{\scriptsize 119a}$,
\AtlasOrcid[0000-0002-1002-1652]{K.~Abeling}$^\textrm{\scriptsize 55}$,
\AtlasOrcid[0000-0001-5763-2760]{N.J.~Abicht}$^\textrm{\scriptsize 49}$,
\AtlasOrcid[0000-0002-8496-9294]{S.H.~Abidi}$^\textrm{\scriptsize 30}$,
\AtlasOrcid[0009-0003-6578-220X]{M.~Aboelela}$^\textrm{\scriptsize 45}$,
\AtlasOrcid[0000-0002-9987-2292]{A.~Aboulhorma}$^\textrm{\scriptsize 36e}$,
\AtlasOrcid[0000-0001-5329-6640]{H.~Abramowicz}$^\textrm{\scriptsize 157}$
% ...
\bigskip
\\

% ...
$^{17}$Department for Physics and Technology, University of Bergen, Bergen; Norway.\\
% ...
$^{30}$Physics Department, Brookhaven National Laboratory, Upton NY; United States of America.\\
% ...
$^{36}$$^{(a)}$Facult\'e des Sciences Ain Chock, Universit\'e Hassan II de Casablanca;$^{(b)}$Facult\'{e} des Sciences, Universit\'{e} Ibn-Tofail, K\'{e}nitra;$^{(c)}$Facult\'e des Sciences Semlalia, Universit\'e Cadi Ayyad, LPHEA-Marrakech;$^{(d)}$LPMR, Facult\'e des Sciences, Universit\'e Mohamed Premier, Oujda;$^{(e)}$Facult\'e des sciences, Universit\'e Mohammed V, Rabat;$^{(f)}$Institute of Applied Physics, Mohammed VI Polytechnic University, Ben Guerir; Morocco.\\
% ...
$^{45}$Physics Department, Southern Methodist University, Dallas TX; United States of America.\\
% ...
$^{49}$Fakult\"{a}t Physik, Technische Universit{\"a}t Dortmund, Dortmund; Germany.\\
% ...
$^{55}$II. Physikalisches Institut, Georg-August-Universit\"{a}t G\"ottingen, G\"ottingen; Germany.\\
% ...
$^{104}$CPPM, Aix-Marseille Universit\'e, CNRS/IN2P3, Marseille; France.\\
% ...
$^{119}$$^{(a)}$New York University Abu Dhabi, Abu Dhabi;$^{(b)}$United Arab Emirates University, Al Ain; United Arab Emirates.\\
% ...
$^{123}$Homer L. Dodge Department of Physics and Astronomy, University of Oklahoma, Norman OK; United States of America.\\
% ...
$^{157}$Raymond and Beverly Sackler School of Physics and Astronomy, Tel Aviv University, Tel Aviv; Israel.\\
% ...
$^{a}$ Also at Affiliated with an institute formerly covered by a cooperation agreement with CERN.\\
$^{b}$ Also at An-Najah National University, Nablus; Palestine.\\
$^{c}$ Also at Borough of Manhattan Community College, City University of New York, New York NY; United States of America.\\
$^{d}$ Also at Center for Interdisciplinary Research and Innovation (CIRI-AUTH), Thessaloniki; Greece.\\
$^{e}$ Also at Centre of Physics of the Universities of Minho and Porto (CF-UM-UP); Portugal.\\
$^{f}$ Also at CERN, Geneva; Switzerland.\\
% ...

\end{flushleft}
